\documentclass[apj]{emulateapj}  
\slugcomment{July 2017} 
\usepackage{dcolumn}
\usepackage{bm}
\usepackage{graphicx}
\usepackage{amssymb,amsmath}
\usepackage{epsfig}
\usepackage{color}
\usepackage{wasysym}
\usepackage{natbib}
\usepackage{twoopt}
\usepackage{dashrule}
\definecolor{slateblue}{rgb}{0.1,0.22,0.58}
\usepackage[breaklinks=true,colorlinks=true,linkcolor=slateblue,citecolor=slateblue,urlcolor=slateblue,pdfauthor={Tomassetti},pdftitle={AbarSnr}]{hyperref}
\def\MyTitle#1{{\section{#1}}} 

\definecolor{ColorTitle}{cmyk}{0,.88,.77,.40}

\newcommand{\AMS}{\textsf{AMS}}
\newcommand{\etal}{et al.}
\newcommand{\eg}{\textit{e.g.}} 
\newcommand{\ie}{\textit{i.e.}}

\newcommand{\Htwo}{\ensuremath{^{2}}\textrm{H}}
\newcommand{\Het}{\ensuremath{^{3}}\textrm{He}}
\newcommand{\Hef}{\ensuremath{^{4}}\textrm{He}}

\renewcommand{\H}{\textrm{H}}
\newcommand{\He}{\textrm{He}}

\newcommand{\BC}{\textrm{B}/\textrm{C}} 

\newcommand{\Li}{\textrm{Li}} 
\newcommand{\Be}{\textrm{Be}} 
\newcommand{\B}{\textrm{B}} 
\newcommand{\C}{\textrm{C}} 
\newcommand{\N}{\textrm{N}} 
\newcommand{\Oxy}{\textrm{O}} 
\newcommand{\Si}{\textrm{Si}} 
\newcommand{\Fe}{\textrm{Fe}} 

\newcommand{\p}{\ensuremath{p}}

\newcommand{\pbarp}{\ensuremath{\bar{p}}/\ensuremath{p}}
\newcommand{\pbar}{{\ensuremath{\bar{p}}}}
\newcommand{\nbar}{{\ensuremath{\bar{n}}}}
\newcommand{\dbar}{\ensuremath{\rm \overline{d}}}

\newcommand{\hebar}{\ensuremath{\rm \overline{He}}}
\newcommand{\htwobar}{\ensuremath{\rm \overline{^{2}H}}}
\newcommand{\hetbar}{\ensuremath{\rm \overline{^{3}He}}}
\newcommand{\htbar}{\ensuremath{\rm \overline{^{3}H}}}
\newcommand{\abar}{\ensuremath{\rm \overline{A}}}

\renewcommand{\S}{\mathcal{S}}

\usepackage{wasysym}
\usepackage{xspace}

\begin{document}
\title{Production and acceleration of antinuclei in supernova shockwaves}
\author{Nicola Tomassetti\,$^{\,\overline{\textsf{d}}}$ and Alberto Oliva\,$^{\,\overline{\textsf{He}}}$}
\address{$^{\overline{\textsf{d}}}$\, Department of Physics and Earths Science, Universit{\`a} di Perugia, and INFN-Perugia, I-06100 Perugia, Italy,}
\address{$^{\overline{\textsf{He}}}$\, Centro de Investigaciones Energ{\'e}ticas, Medioambientales y Tecnol{\'o}gicas CIEMAT, E-28040 Madrid, Spain,}


\begin{abstract}
  We compute the energy spectra of antideuterons (\dbar) and antihelium (\hebar) in cosmic rays (CRs) in a scenario
  where hadronic interactions inside supernova remnants (SNRs) can produce a diffusively shock-accelerated ``source component'' of secondary antinuclei.
  The key parameters that specify the SNR environment and the interstellar CR transport are tightly constrained
  with the new measurements provided by the \AMS{} experiment on the \BC{} ratio and on the \pbarp{} ratio.
  The best-fit models obtained from the two ratios are found to be inconsistent with each other,
  as the \pbarp{} data require enhanced secondary production. 
  Thus, we derive conservative (\ie, \BC{}-driven) and speculative (\pbarp{}-driven) upper limits to the
  SNR flux contributions for the \dbar{} and \hebar{} spectra spectra in CRs, 
  along with their standard secondary component expected from CR collisions in the interstellar gas.
  We find that the source component of antinuclei can be appreciable at kinetic energies above a few $\sim$10\,GeV/n,
  but it is always sub-dominant below a few GeV/n, that is the energy window where dark-matter (DM) 
  annihilation signatures are expected to exceed the level of secondary production. 
  We also find that the \emph{total} (standard + SNR) flux of secondary \dbar{} and \hebar{}
  is tightly constrained by the data. Thus, the presence of interaction processes in SNRs does not
  critically affect the total background for DM searches.
\end{abstract}
\keywords{cosmic rays --- acceleration of particles --- ISM: supernova remnants}
\maketitle

\MyTitle{Introduction}  
%
The observation of antinuclei in the cosmic-ray (CR) flux may provide a unique 
discovery avenue to the particle nature of cosmological dark matter (DM).
Annihilation or decay processes of DM particles in the Galaxy may generate antiproton (\pbar) and antineutron (\nbar) particles
that, in turn, can merge into heavier antinuclei such as antideuteron \htwobar{} (or \dbar) or antihelium \hetbar{} (in short, \hebar). 
These particles are excellent DM messengers for a wide range of masses where the expected signals lie 
in the energy window below a few GeV. 
At these energies, the level of secondary production is expected to be suppressed,
thanks to the kinematics of antinuclei production from CR collisions with the cold gas
and to the power-law falling shape of the colliding CR spectra \citep{Aramaki2016,IbarraWild2013,Dal2015,Cirelli2014,Carlson2014}. 
In contrast to antiprotons, \dbar{} or \hebar{} nuclei have never been observed in the cosmic radiation, 
but a long series of CR detection experiments has established tight upper limits to the flux of these particles.
Very recently, hints for \hebar{} events may have been identified by \AMS, 
rising the possibility of an excess of nuclear antimatter in CRs \citep{Sokol2017,Coogan2017,Blum2017}. 

In this \emph{Letter}, we report calculations of CR nuclei and antinuclei energy spectra in a model of CR acceleration
and transport where the production of secondary particles can also take place inside Galactic sources. 
Our calculations are carried out in the framework of the linear diffusive-shock-acceleration (DSA) theory and 
the two-halo model of diffusive propagation.
Statistical analyses
based on standard $\chi^{2}$-techniques
are performed using new \AMS{} data on boron-to-carbon (\BC) and antiproton-to-proton (\pbarp) ratios.
While the \BC{} ratio is found to decrease steadily up to 1\,TeV/n of kinetic energy per nucleon,
the measured \pbarp{} ratio appears to be remarkably constant at  $E\gtrsim$\,60\,GeV, thus suggesting an enhanced antiproton production from supernova remnants (SNRs).
Using both ratios separately in order to calibrate the secondary production, we present conservative
(\ie, \BC{}-driven) and speculative (\pbarp{}-driven) predictions for the \dbar{} and \hebar{} fluxes in CRs.
We then discuss our results in the context of DM searches in space.
\\

\MyTitle{Calculations}    
\label{Sec::Calculations} 
%
Our goal is to compute the spectrum of CR nuclei and antinuclei near Earth. 
This involves three steps:
(i) nuclear-physics calculations for the production of CR nuclei and antinuclei from CR+gas collisions;
(ii) DSA-based calculations of primary and secondary CR spectra injected in the ISM by SNRs;
(iii) CR propagation calculations including secondary production, Galactic transport, and solar modulation \citep[for reviews, see][]{Grenier2015,AmatoBlasi2017}.

These calculations involve a large number of reactions $k+i\rightarrow j$, describing the generation
of $j$-type particles from fragmentation of $k$-type CR nuclei off $i$-type target atoms in the ISM or in the SNR background plasma.
We account for the production of several isotopes such as  \Htwo, \Het, $^{6,7}$\Li, $^{7,9,10}$\Be, and $^{10,11}$\B{} from the disintegration
of heavier nuclei, and in particular \C-\N-\Oxy, \Si, and \Fe.
The adopted cross-section are those re-evaluated in earlier work \citep{Tomassetti2012Iso,Tomassetti2015XS}. 
For these species, cross-sections are expressed as energy-dependent functions $\sigma_{k \rightarrow j}^{i}(E)$,
thanks to the \emph{straight-ahead} approximation.
The production of antiparticles requires double-differential cross-sections.
For antinucleons (\pbar{} and \nbar) we have implemented the algorithm of \citet{DiMauro2014} (from Eq.13), which gives the Lorentz-invariant
distribution function $f^{p\rightarrow\bar{p}}_{p} \equiv E_{\bar{p}}\frac{d^{3}\sigma}{dp^{3}_{\bar{p}}}$ for \p-\p{} collision processes.
For other CR+ISM collisions (\p-\He, \He-\p, \He-\He) we used dedicated parameterizations \citep{InelasticXS}. 
Heavier nuclei \dbar{} and \hebar{} are formed from the fusion of \pbar{} and \nbar{} in hadronic jets of nucleus-nucleus collisions. 
Their production cross-sections are evaluated within the framework of the nuclear coalescence model. 
According to this model, an antinucleus is formed when the relative momenta of all pairs of antinucleons
lie within the so-called coalescence momentum $p_{0}$, which is a free parameter. 
The formation of $\bar{A}$ antinuclei from $Z$ antiprotons and $A-Z$ antineutrons is calculated from
\begin{equation} \label{Eq::Coalescence}
 E_{\bar{A}} \frac{d^{3}N_{\bar{N}}}{dp^{3}_{\bar{A}}} = 
B_{\bar{A}} \times \left( E_{p} \frac{ d^{3}N_{p}}{dp^{3}_{p}} \right)^{Z}  \times \left( E_{n} \frac{ d^{3}N_{n}}{dp^{3}_{n}} \right)^{A-Z}\,,
\end{equation}
where $N_{\bar{A}}$, $N_{p}$, and $N_{n}$ are the production multiplicities. 
The coefficient $B_{\bar{A}}$ is linked to the probability of having $Z$ \pbar{} and $A-Z$ \nbar{} within distance $p_{0}$ in momentum space,
$B_{\bar{A}}\propto \left(\frac{4\pi}{3}p_{0}^{3}\right)^{A-1}\frac{m_{\bar{A}}}{m^{A}_{p}}$,
where dedicated near-threshold corrections are applied as in \citet{Chardonnet1997}.
We take the coalescence momentum $p_{0}\cong 90$\,MeV/c, based on data of the ISR accelerator at CERN \citep{Alper1973,Gibson1978}.
Note that in Eq.\,\ref{Eq::Coalescence} 
we have modified the coalescence scheme in order to account for asymmetric \nbar/\pbar{} production.
Following \citet{DiMauro2014},  we set a $30\,\%$ asymmetry between \nbar{} and \pbar{} cross-sections.
This leads to a larger production of \htbar{} triplets (\pbar,\nbar,\nbar) with respect to \hetbar{} triplets (\pbar,\pbar,\nbar).
Because the former decays rapidly into \hebar, the near-Earth \hebar{} flux  includes both contributions.
Finally, we account for destruction processes of these particles and for their subsequent production of tertiaries,
including non-annihilating reactions such as \abar+\p+\abar$^{\prime}+X$ \citep{InelasticXS,Donato2008,Duperray2005}.
Further details on nuclear physics calculations will be presented in future work (in preparation).

The spectrum of CRs accelerated in SNRs is calculated using the linear DSA theory,
where we account for production and destruction processes of secondary and tertiary fragments.
Similar calculations are done in earlier work for CR nuclei \citep{MertschSarkar2009,MertschSarkar2014,TomassettiDonato2012} 
and antiparticles \citep{BlasiSerpico2009,Kohri2016,Herms2017} and leptons \citep{Blasi2009,Serpico2012,TomassettiDonato2015}.
We closely follow the derivation of \citet{TomassettiFeng2017}. 
We considered primary nuclei \p, \Hef, \C, \N, \Oxy, \Si, and \Fe{} tuned to recent data \citep{MertschSarkar2014},
and secondary production of \H-\He-\Li-\Be-\B{} isotopes and antinuclei.
For antinuclei, we drop the ``inelasticity approximation'' that links the antiproton momentum $p$ to the primary proton momentum $p_{p}$ \citep{Herms2017}.
It has been demonstrated that this assumption leads to an overestimation of the secondary production \citep{Kachelriess2011}.
For each species the CR flux injected by SNRs in the ISM 
is evaluated by integrating all solutions over the SNR acceleration history
and accounting for an average SNR explosion rate per unit volume in the Galaxy.
For secondary particles,
the resulting source term $\S^{\rm snr}(R)$  as a function of rigidity $R=p/Z$
is made up of two components called $\mathcal{A}$ and $\mathcal{B}$ \citep{Kachelriess2011,BlasiSerpico2009}.
The $\mathcal{B}$--term describes secondary CRs which, after production, are advected downstream without experiencing further DSA.
Their spectrum is similar of that of standard primary CRs $\S^{\rm snr}_{B} \sim R^{-\nu}$ with $\nu\approx$\,2.
The $\mathcal{A}$--term describes CRs produced close the shock that are still able to start DSA,
reaching a distribution of the type $\S_{A}^{\rm snr} \sim R^{-\nu+1}$, \ie, harder by one in the power law for an assumed Bohm-like
diffusion coefficient $D \equiv \kappa D_{B} \approx 3\kappa 10^{22}R_{\rm GV}/B_{\mu\,G}$\,cm$^{2}$/s.
The amplitudes of both terms scale with on the product of SNR age, $\tau_{\rm snr}\cong$\,20\,kyr,
and the density of the unshocked plasma $n_{-}$, which is of the order of 1\,cm$^{-3}$.
The $\mathcal{A}$--term depends also on the ratio $D/u_{-}^{2}$
between diffusion coefficient and shockwave speed. 
Here, we fix $u_{-}\cong\,5\cdot$10$^{7}$\,cm/s and use the $\kappa/B$ ratio as free parameter.

\begin{table*}[!ht]
\caption{\small
  Fit results for the two propagation scenarios and for various minimal values of kinetic energy.
  \label{Tab::FitSummary}
}
\centering
\begin{tabular}{cc c cccc c cccc} \hline\hline
\\[-2mm]
\multicolumn{2}{c}{Data} & {} &  \multicolumn{4}{c}{ B/C ratio } &   {} & \multicolumn{4}{c}{ $\bar{p}/p$ ratio } 
\\[1mm]
\cline{1-2}\cline{4-7}\cline{9-12}
\\[-2mm]
  {Model of CR} & {$E_{\rm min}$}   & {}  & {$K_{0}/L$} & $\kappa/B$ & $\tau_{\rm snr} n_{-}$ & {$\chi^{2}/df$} & {} & {$K_{0}/L$} & $\kappa/B$ & $\tau_{\rm snr} n_{-}$ & {$\chi^{2}/df$}\\ 
  {propagation} & {(GeV/n)}   & {}  & {(kpc/Myr)} & ($\mu G^{-1}$) & (kyr/cm$^{3}$) & {\dots} & {} & {(kpc/Myr)} & ($\mu G^{-1})$ & (kyr/cm$^{3}$) & {\dots}
  \\[1mm]
\hline
\\[-2mm]
Kolmogorov & 10  & {} &  
0.0165 $\pm$ 0.0003 & 0.0 $\pm$ 5.8 & 24.0 $\pm$ 2.0 & 14/38 & {} & 
0.0165 $\pm$ 0.0011 & 9.8 $\pm$ 2.9 & 46.5 $\pm$ 8.1 & 11/31 
\\
Kraichnan\, & 10  & {} &  
0.0176 $\pm$ 0.0003 & 0.01 $\pm$ 0.13 & 45.6 $\pm$ 1.6 & 21/38 & {} & 
0.0152 $\pm$ 0.0001 & 6.45 $\pm$ 2.46 & 59.0 $\pm$ 5.0 & 12/31
\\
Kolmogorov & 3  & {} &  
0.0159 $\pm$ 0.0002 & 0.0 $\pm$ 0.3 & 27.1 $\pm$ 1.6 & 22/50 & {} &
0.0134 $\pm$ 0.0006 & 4.3 $\pm$ 2.4 & 67.4 $\pm$ 4.9 & 25/43
\\
Kraichnan\, & 3  & {} &  
0.0155 $\pm$ 0.0002 & 0.01 $\pm$ 0.06 & 52.6 $\pm$ 1.2 & 77/50 & {} &
0.0111 $\pm$ 0.0003 & 0.5 $\pm$ 32   & 76.5 $\pm$ 1.6 & 36/43
\\
Kolmogorov & 1  & {} &  
0.0156 $\pm$ 0.0002 & 0.01 $\pm$ 0.18 & 29.1 $\pm$ 1.4 & 31/59 & {} &
0.0129 $\pm$ 0.0005 & 3.19 $\pm$ 2.26 & 70.9 $\pm$ 4.2 & 27/50
\\
Kraichnan\, & 1  & {} &  
0.0148 $\pm$ 0.0002 & 0.01 $\pm$ 0.04 & 55.9 $\pm$ 1.1 & 122/59 & {} &
0.0105 $\pm$ 0.0003 & 0.2 $\pm$ 0.86 & 78.8 $\pm$ 1.4 & 46/50
\\[1mm]
\hline\hline
\end{tabular}
\end{table*}
%
\begin{figure*}[!t]
\epsscale{1.15}
\plotone{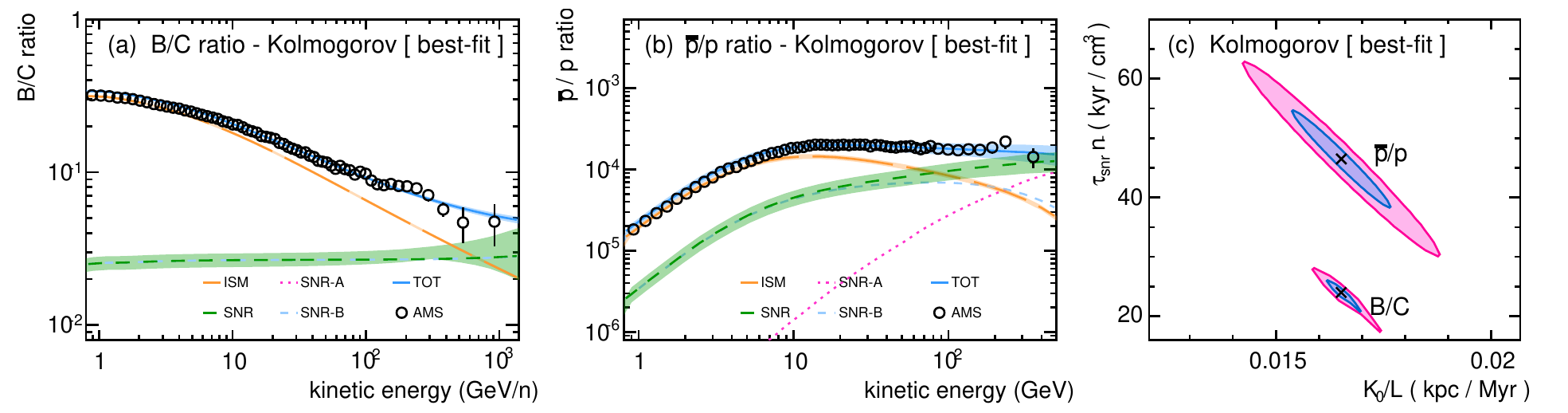}
\caption{\footnotesize%
  \BC{} ratio (a) and \pbarp{} ratio (b) measured by \AMS{} \citep{Aguilar2016PbarP,Aguilar2016BC}
  in comparison with the best-fit models (light blue solid lines) with $E_{\rm min}=$\,10\,GeV/n of kinetic energy.
  Contributing components from standard secondary production (orange, long-dashed lines) are shown together with the
  component from secondary production in SNRs (green short-dashed lines). The shaded bands represent the errors from the fits.
  Two-dimensional contour plots (c) are shown for the parameters $\tau_{\rm snr}n_{-}$ and $K_{0}/L$
  at 68\,\% (shaded blue area) and 95\,\% (shaded pink area) of confidence levels.
  The plots correspond to \BC-driven (a) and \pbarp-driven (b) fits performed at $E>10$\,GeV.
}
\label{Fig::ccSecPriRatiosFits}
\end{figure*}

To model the subsequent propagation of CRs in the ISM, we adopt a two-halo model of
CR diffusion and nuclear interactions \citep{Tomassetti2012Hardening}.
The Galaxy is modeled as a disk of half-thickness $h\cong$\,100\,pc 
containing SNRs and gas with number density $\tilde{n}\cong$\,1\,cm$^{-3}$. 
The disk is surrounded by a diffusive halo of total half-thickness $L$ and zero matter density. 
The diffusion coefficient is taken as $K(R,z) = \beta K_{0}(R/GV)^{\hat{\delta}(z)}$ for all particles,
where $R=p/Z$ is the particle rigidity (momentum/charge ratio), and $K_{0}$ expresses its normalization.
Its $z$-dependence is accounted for in the scaling index $\hat{\delta}(z)$.
We set up the two halos using $\hat{\delta}=\delta_{0}$ in the region of $|z|<\xi\,L$ (close to the disk) and  $\hat{\delta}=\delta_{0}+\Delta$
for $|z|>\xi\,L$ (away from the disk), with $\xi\cong 0.1$, $L\cong$\,5\,kpc, and $\Delta=0.55$ \citep{Tomassetti2015TwoHalo,Feng2016}.

In this model, the $\delta_{0}$ parameter is linked to the spectrum of pre-existing (SNR-driven) turbulence near the Galactic disk.
Away from the disk the spectral index is $\delta_{0}+\Delta$, reflecting the effects of CR-driven turbulent motion. 
The $\delta_{0}$ parameter determines the secondary production at high energy. We consider two distinct scenarios, based on
Kolmogorov-type turbulence in the proximity of the Galactic disk ($\delta_{0}=1/3$), and based on a Iroshnikov-Kraichnan
turbulence spectrum ($\delta_{0}=1/2$), respectively. The total source term contains a DSA/SNR injection term, $\S^{\rm snr}$, 
and a term describing secondary production in the ISM, $\S^{\rm sec}= \sum_{k} \tilde{\Gamma}_{k}^{\rm fr} \mathcal{N}_{k}$.
To compute the fragmentation rates in the ISM $\tilde{\Gamma}^{\rm in/fr}$ we adopt
the same nuclear network (and the same cross-sections) as occurring inside sources. 

Solar modulation is calculated under the force-field approximation, using the so-called
\emph{modulation potential} parameter $\phi\cong$\,0.7\,GV for the \AMS{} observation period \citep{Ghelfi2017}.
This value has been cross-checked using CR proton data \citep{Aguilar2015Proton}. 
For antinuclei, the modulation is in general different due to charge-sign-dependent effect arising from drift motion.
We have investigated this difference \emph{a posteriori} by means of a numerical 2D model that accounts for particle drift. 
We found no appreciable charge-sign difference for the \AMS{} data because these data
have been collected during a period including the polarity reversal of early 2013,
and thus they contain samples of CR particles propagated under both polarity conditions.
Hence, we have adopted the same modulation potential for all CR particles.

The new \AMS{} data on the \BC{} ratio \citep{Aguilar2016BC} and on the \pbarp{} ratio \citep{Aguilar2016PbarP}
have been used to constrain three key parameters that are directly linked to physical observables:
(i) the ratio $K_{0}/L$ between the diffusion coefficient normalization and the half-height of the halo,
which governs the secondary production in the ISM ($N_{\rm s} \sim N_{\rm p}\times L/K\times\Gamma^{\rm fr}_{s\rightarrow p}$);
(ii) the product $\tau_{\rm snr}n_{-}$ between SNR age and upstream plasma density,
which regulates the secondary production inside SNRs ($N_{\rm s}\propto N_{\rm p}\tau_{\rm snr}n_{-}$);
and (iii) the $\kappa/B$ ratio, which set the normalization of the diffusion coefficient $D$ at the shock, 
giving the yield of DSA-accelerated secondaries ($N_{\rm s}\propto N_{\rm p} D/u^{2}_{\pm} \propto \kappa/B$). 
These parameters are constrained by means of a standard $\chi^{2}$ analysis where the \BC{} and \pbarp{} data from \AMS{} are used separately,
above the kinetic energy $E_{\rm min}$=1, 3, and 10\,GeV.
The use of larger $E_{\rm min}$ values gives more
reliable results because the low-energy region can be affected by uncertainties in solar modulation.
We consider two model implementations: \emph{conservative}, \ie, \BC-driven parameter constraints, and \emph{speculative}, \ie,
\pbarp-driven parameter constraints.
\\

\begin{figure*}[!t]
\epsscale{0.96}
\plotone{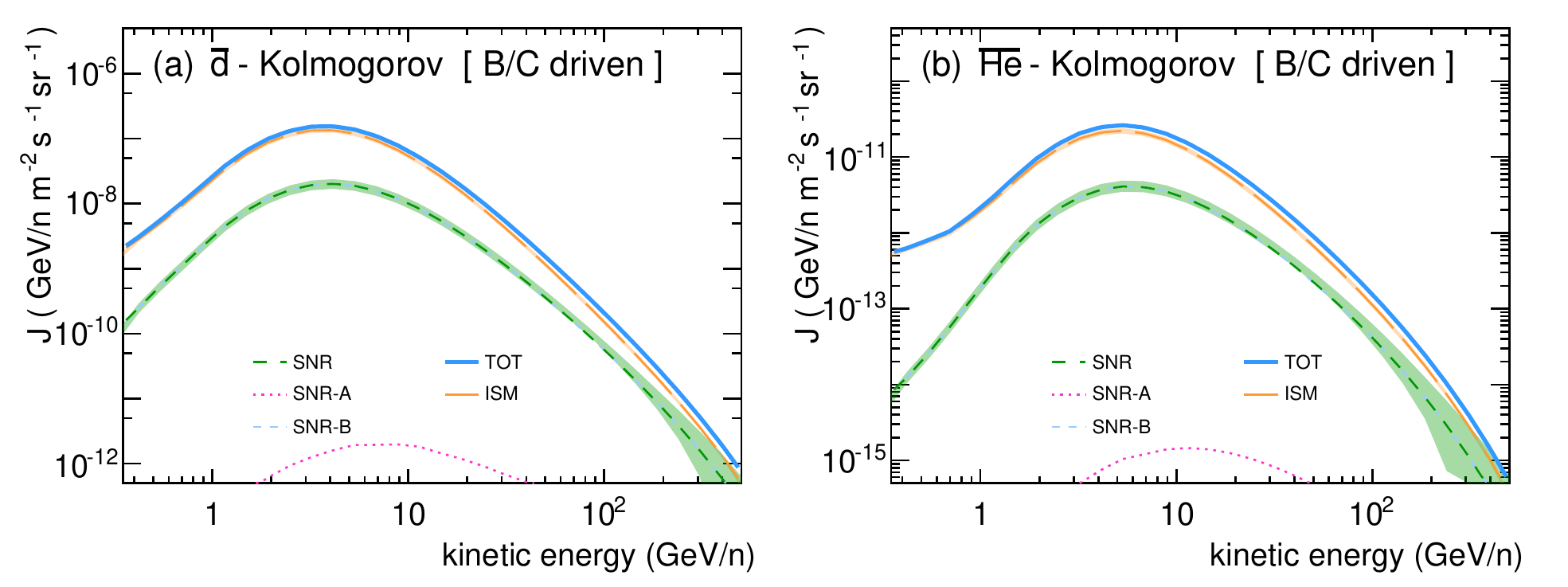}
\qquad
\qquad
\qquad
\plotone{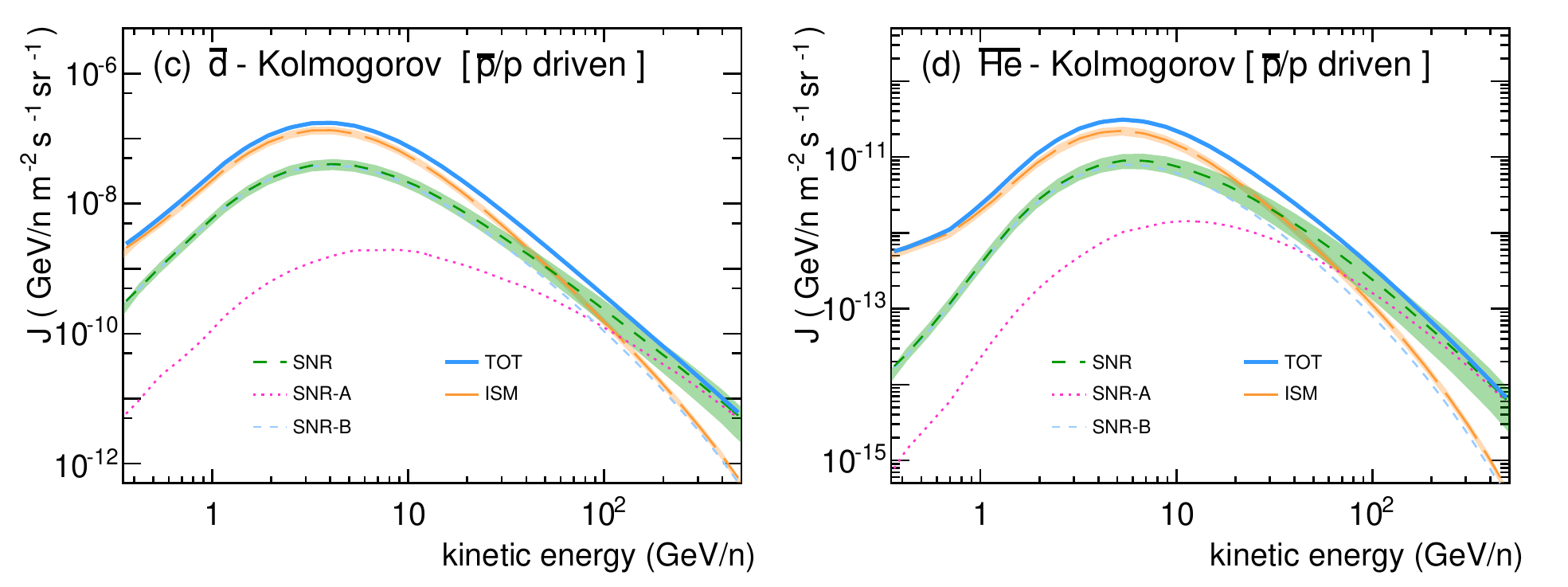}
\caption{
  Model predictions of for the total flux of antideuterons (left) and antihelium (right), including secondary production inside SNRs 
  (green short-dashed lines), standard production in the ISM (orange long-dashed lines), and total flux (light blue solid lines).
  Calculations are shown from the \BC-driven (top) and \pbarp-driven (bottom) fit at kinetic energy $E>10$\,GeV/n under the Kolmogorov model. 
}\label{Fig::ccAbarBCDriven}\label{Fig::ccAbarPbarPDriven}\label{Fig::ccAbar}
\end{figure*}

\MyTitle{Results and Discussion}  
%
The fit results are summarized in Table\,\ref{Tab::FitSummary}. 
To keep the results concise, we will focus on results obtained with $E_{\rm min}=10$\,GeV/n under the Kolmogorov model.
The best-fit models are shown in Fig.\,\ref{Fig::ccSecPriRatiosFits} for the \BC{} (a) and \pbarp{} (b) driven fits.
Overall, we found that all configurations lead to good fits for both observables.
Accounting for secondary production processes inside SNRs improves the description of the \AMS{} data
because, at high-energy, the expected decrease of the ISM-induced \BC{} ratio is balanced by the harder SNR-accelerated boron flux.
In Kraichnan-like models, even larger SNR fluxes are inferred because under this scenario the ISM-induced \BC{} ratio is steeper.
However, the Kolmogorov-type model appears to be favored. This model is also supported by recent observations of
magnetic turbulence in the local ISM \citep{Burlaga2015}. 
Interestingly, the inferred SNR gas density is found to be consistent with the
average density of the ISM ($\sim$\,1\,cm$^{-3}$), as one would expect from a large population of contributing SNRs.
The appearance of the SNR component does not give significant signatures on the \BC{} ratio, but
it causes a progressive \BC-hardening that introduces new degeneracies between source and transport parameters.
As noted in  \citet{TomassettiDonato2012}, disregarding interactions in SNRs may leads
to biased results for the $\delta_{0}$ parameter. 
This effect may explain why in \citet{Feng2016}, 
from the \BC{} data, it was inferred $\delta_{0}\approx$\,0.18$\pm$0.1 in terms of MCMC posterior mean.
It is interesting to note that in \citet{Aloisio2015}, under a phenomenologically similar scenario of CR propagation,
an SNR grammage of $X_{\rm snr}\approx$\,0.16\,g\,cm$^{-2}$ was invoked to fit the \BC{} data.
This grammage corresponds to $\tau_{\rm snr}\approx$\,20\,kyr and $n_{-}\approx$\,1\,cm$^{-3}$, in good agreement with our findings.

From Fig.\,\ref{Fig::ccSecPriRatiosFits}(b), it can be noted that \pbarp-driven fits lead to stronger SNR production,
recovering the recent results of \citet{Cholis2017}.
However, it can be seen that these results are inconsistent with those obtained with the \BC{} ratio.
In Fig.\,\ref{Fig::ccSecPriRatiosFits}(c), 68\,\% and 95\,\% probability contours are shown for 
the key parameters $K_{0}/L$ and $\tau_{\rm snr}n_{-}$. 
The best fits associated with the two ratios lie in separate regions of the parameter space.
Even larger discrepancies are found for the $\kappa/B$ parameter, giving rise to the $\mathcal{A}$--term of Fig.\,\ref{Fig::ccSecPriRatiosFits}(b).
The interpretation of the \pbarp{} data in terms of \pbar{} acceleration in SNRs
requires dense background media and exceedingly fast diffusion, suggesting strong magnetic damping.
Such a peculiar SNR environment may be in conflict with the usual requirements of having efficient DSA 
up to $R\sim$\,10\,TV or above \citep[see discussions in][]{Kachelriess2011,Serpico2012}. 
In this respect, \BC-driven results lead to more plausible properties for CR accelerators.
A possible source of error may be linked to
\pbar{} production cross-sections. For instance, different calculations \citep[\eg,][]{Kachelriess2015,Winkler2017,Feng2016}
may give larger \pbar{} production, therefore leading to smaller SNR fluxes and better agreement with the \BC-driven results.

In Fig.\,\ref{Fig::ccAbar}, we present calculations for the expected fluxes of \dbar{} and \hebar{} nuclei
under the Kolmogorov model. Model predictions are shown from conservative (\BC{}-driven) and speculative (\pbarp{}-driven) parameter constraints.
Colors and line styles are encoded as in Fig.\,\ref{Fig::ccSecPriRatiosFits}.
Shapes and the intensities of these fluxes are in agreement with those reported in earlier works \citep{Aramaki2016,Cirelli2014}.
All fluxes show a characteristic peak at a few GeV/n of kinetic energy
preceded and followed by quick spectral drops at lower and higher energies.
These drops reflect the kinematics of antinuclei production and the rapid power-law falling flux of the progenitors, respectively.
From the \BC-driven predictions of Fig.\,\ref{Fig::ccAbar}(a) and Fig.\,\ref{Fig::ccAbar}(b),
we have found that the SNR flux contribution is sub-dominant 
by one order of magnitude in the considered energy range 0.5\,--\,100 GeV/n, and
in particular the $\mathcal{A}$--term is highly suppressed.
In contrast,
from \pbarp-driven predictions of Fig.\,\ref{Fig::ccAbar}(c) and Fig.\,\ref{Fig::ccAbar}(d),
an enhanced flux of SNR-accelerated antinuclei is predicted. This SNR component dominates the total flux at $E\gtrsim$\,30\,GeV/n.
When approaching $E\sim$\,500\,GeV/n, the total fluxes of \dbar{} and \hebar{}
are one order of magnitude larger than those arising from \BC-driven calculations. 
In the high-energy region, however, the flux intensities are
experimentally inaccessible by the existing or planned projects of CR detection.
Interestingly, in the sub-GeV/n energy region, \pbarp-driven calculations do not show substantial differences from the \BC-driven results.
At these energies, secondary production processes in SNRs do not provoke significant modifications on the expected fluxes.
Similar conclusions can be drawn within the Iroshnikov-Kraichnan model.
In fact, we emphasize that the spectra of secondary antinuclei suffer from the same types of parameter degeneracy as \BC{} and \pbarp{} ratios.
As results, in spite of uncertainties on the amplitude of SNR and ISM components,
the \emph{total} ISM+SNR flux prediction is rather stable for a large region of parameter space.
Moreover, \BC- and \pbarp-driven predictions give similar results for the total flux intensity
in the GeV/n energy region, even though the estimated SNR and ISM contributions are substantially different.
In this respect, the
expected level of background for DM searches in the low-energy region
appears to be
hardly affected by interaction processes inside SNRs.
We expect that these effects do not apply to the shape of fluxes coming from DM annihilation.
The DM-annihilation-induced fluxes suffer from different types of parameter degeneracy
because the DM source is distributed in the whole Galactic halo. Thus, when modeling DM-induced CR particles,
precision data on the \BC{} ratio are not sufficient to fully characterize their interstellar transport.
Finally, we stress once more that calculations of CR antinuclei are affected by sizable cross-section-induced uncertainties.
For antideuterons, laboratory \p-\p{} data give relatively good constraints to their production rate.
For antihelium the situation is more uncertain. In a recent estimate \citep{Blum2017}, 
the \hetbar{} and \htbar{} production cross-sections are found to be nearly 50 times larger than those used in our work.
These cross-sections have a similar influence on both ISM and SNR components. 
\\

\MyTitle{Conclusions}  
%
The discovery of antinuclei in the cosmic radiation could be the next milestone in CR physics.
With unmatched performance, the  sensitivity of the \AMS{} experiment is gradually approaching 
the expected level of secondary production in the ISM \citep{Kounine2011}.
While this search is now in progress, hints for \hebar{} events may have been identified \citep{Sokol2017,Coogan2017,Blum2017},
Meanwhile, the first science flight of the GAPS antimatter detection project has been approved by NASA  \citep{GAPS}.
Hence, we believe, perhaps optimistically, that the first observation of CR antinuclei
has concrete chances to be achieved by the present or incoming generation of CR detection experiments.

In this work, we have used new measurements of secondary-to-primary ratios reported by \AMS{}
to determine the flux of secondary antinuclei produced and accelerated in SNR shockwaves.
As we have shown, the SNR flux component of CR antinuclei can be appreciably large at high energy,
especially if the model is calibrated using data on the \pbarp{} ratio.
However, in the low-energy window between $\sim$\,0.1 and a few GeV/n, which is where DM-induced signatures
have chances to exceed over the background, this component is found to be always sub-dominant.
Furthermore, we found that the \emph{total} ISM+SNR flux of antinuclei
is rather stable for a large region of parameter configurations, and for both
scenarios of CR propagation considered, even though the single SNR and ISM induced contributions are
different and highly model dependent.
Thus, the estimated level of secondary production in the low-energy window,
\ie, the background for DM searches,
appears to be only barely affected by interaction processes inside SNRs.
\\[0.12cm]
{\footnotesize%
We thank our colleagues of the AMS Collaboration for valuable discussions.
AO acknowledges CIEMAT, CDTI and SEIDI MINECO under grants ESP2015-71662-C2-(1-P) and MDM-2015-0509.
NT acknowledges support from MAtISSE.
This project has received funding from the European Union's Horizon 2020 research and innovation programme under the Marie Sklodowska-Curie grant agreement No 707543.
}
\\


\end{document}